\documentstyle[preprint,aps,eqsecnum]{revtex}
\tighten
\begin{document}
\preprint{LPTHE-Paris-94-21/DEMIRM-Paris-94014}
\title{\bf BACK REACTION OF STRINGS IN \\
 SELF CONSISTENT STRING COSMOLOGY}
\author{{\bf H.J. de Vega$^{(a)}$
 and N. S\'anchez$^{(b)}$}}
\address
{  (a)  Laboratoire de Physique Th\'eorique et Hautes Energies,
Universit\'e Pierre et Marie Curie (Paris VI) et Universit\'e Denis
Diderot (Paris VII),
Tour 16, 1er. \'etage, 4, Place Jussieu
75252 Paris, cedex 05, France. \\ Laboratoire Associ\'{e} au CNRS URA280.\\
 (b) Observatoire de Paris, DEMIRM, 61, Avenue de l'Observatoire,
75014 Paris, France.  Laboratoire Associ\'{e} au CNRS URA336,
Observatoire de Paris et \'{E}cole Normale Sup\'{e}rieure.}
\date{May 1994}
\maketitle
\begin{abstract}
We compute the string energy-momentum tensor and {\bf derive} the
string equation of state from exact string dynamics in cosmological
spacetimes. $1+1,~2+1$ and $D$-dimensional universes are treated for
any expansion factor $R$. Strings obey the perfect fluid relation $ p
= (\gamma -1) \rho $ with three different behaviours: (i) {\it Unstable}
for $ R \to \infty $ with growing energy density $ \rho \sim R^{2-D} $,
{\bf negative} pressure, and $ \gamma =(D-2)/(D-1) $; (ii)
  {\it Dual} for $ R \to
0 $, with $ \rho \sim R^{-D} $, {\bf positive} pressure and
$\gamma = D/(D-1) $ (as radiation); (iii) {\it  Stable} for  $ R \to \infty $
with  $ \rho \sim R^{1-D} $,  {\bf vanishing}  pressure and  $\gamma = 1 $
(as cold matter). We find the back reaction effect of these strings on
the spacetime and we take into account the quantum string decay
through string splitting. This
is achieved by considering {\bf self-consistently} the strings as
matter sources for the Einstein equations, as well as for the
complete effective string equations. String splitting exponentially
suppress the density of unstable strings for large $R$. The
self-consistent solution to the Einstein equations for string
dominated universes exhibits the realistic matter dominated behaviour
$ R \sim (X^0)^{2/(D-1)}\; $ for large times and the radiation dominated
 behaviour $ R \sim (X^0)^{2/D}\; $ for early times.
De Sitter universe  does not emerge as solution of
the effective string equations. The effective
string action (whatever be the dilaton, its potential and the
central charge term) is not the appropriate framework in which to
address the question of string driven inflation.
\end{abstract}

\newpage

\section{\bf Introduction and Results}

Recently, several interesting progresses in the understanding of string
propagation in cosmological spacetimes have been made
\cite{prd}-\cite{ijm}. The classical string equations of motion plus
the string constraints were shown to be exactly integrable in
D-dimensional de Sitter spacetime, and equivalent to a
Toda-type  model with a potential unbounded from below. In 2+1
dimensions, the string dynamics in de Sitter spacetime is exactly
described by the sinh-Gordon equation.

{\bf Exact} string solutions were systematically found by soliton
methods using the linear system associated to the problem
(the so-called dressing method in soliton theory)
\cite{dms,cdms}. In addition, exact circular string solutions
were found in terms of elliptic functions\cite{dls}. All these solutions
describe one string, several strings or even an infinite number of
different and independent strings. A single world-sheet simultaneously
describes many different strings. This is a new feature appearing as a
consequence of the interaction of the strings with the spacetime geometry.
Here, interaction among the strings (like splitting and merging) is neglected,
the only interaction is with the curved background. Different types of
behaviour appear in the multistring solutions. For some of them the
energy and proper size are bounded (`stable strings') while for many
others the energy and size blow up for large radius of the universe
($R \to \infty$ ) (`unstable strings'). In
addition, such stable and  unstable string behaviours are exhibited by
the ring solutions found in ref.\cite{din} for
Friedmann-Robertson-Walker (FRW) universes and for power type inflationary
backgrounds. In all these works, strings were considered as {\it test}
objects propagating on the given {\it fixed} backgrounds.

In the present paper we go further in the investigation of the
physical properties of the string solutions above mentioned. We
compute the energy-momentum tensor of these strings and we use it to
find the back reaction effect on the spacetime. That is, we investigate
whether these classical strings can  sustain the corresponding cosmological
background. This is achieved by considering {\bf self-consistently},
the strings as matter sources for the Einstein (general relativity)
equations (without the dilaton field), as well as for the
string effective equations (beta functions) including the dilaton, the
dilaton potential and the central charge term.

 In spatially homogeneous and isotropic universes,
\begin{equation}
ds^2 = (dX^0)^2 - R(X^0)^2 \sum_{i=1}^{D-1}(dX^i)^2
\label{metA}
\end{equation}
the string energy-momentum tensor $T_A^B(X) \; , (A,B=1,\ldots D) $
for our string solutions takes the fluid form, allowing us to define
the string pressure $p$ through $\, -\delta_i^k~p = T_i^k \, $ and the string
energy density as $ \rho = T_0^0 $. The continuity equation
$D^A\;T_A^B = 0 $ takes then  the form
\begin{equation}
\dot{\rho} + (D-1)\, H\, (p + \rho ) = 0 ,
 \label{contA}
\end{equation}
where
$ H  \equiv  {{d\log R}\over{dX^0}} $.
We consider $D=1+1, D=2+1$, and generic $D-$dimensional universes.

In $1+1$ cosmological spacetimes we find the general solution of the
string equations of motion and constraints for arbitrary expansion
factor $ R $ . It consists of two families : one depends on two
arbitrary functions $ f_{\pm}(\sigma \pm \tau) $ and has {\it constant}
energy density $ \rho $ and  {\it negative} pressure $
p = -  \rho $. That is, a perfect fluid relation holds
\begin{equation}
p = (\gamma - 1 ) \rho
\label{flup}
\end{equation}
with $\gamma = 0$ in $ D = 1 + 1 $ dimensions. The other family of
solutions depends on {\it two} arbitrary constants and describes a
massless point particle (the string center of mass). This second
solution has $  p = \rho =  u \; R^{-2} > 0 $ . This is a perfect
fluid type relation with $\gamma = 2 $.
These behaviours fulfil the continuity equation (\ref{contA}) in $ D = 2$.

In $2+1$ dimensions and for any factor $ R $ , we find that circular
strings exhibit three different asymptotic behaviours :
\begin{itemize}
\item (i) {\bf unstable} behaviour for $ R \to \infty $  (this
corresponds to conformal time $ \eta \sim \tau \to \tau_0 $ with
 finite $ \tau_0 $ and proper
string size $ S \sim R \to \infty $), for which the string energy $ E_u
\sim R \to \infty $ and the string pressure $ p_u \simeq -E_u/2 \to -\infty
$ is {\bf negative} .
\item (ii) {\bf Dual} to unstable behaviour for $ R \to 0 $. This
corresponds to  $ \eta \sim (\tau - \tau_0)^{-1} \to +\infty $ for
finite $ \tau  \to \tau_0 $, $ S \sim R \to 0 $
(except for de Sitter spacetime
where $ S \to 1/H $ ), for which the string energy $ E_d \sim 1/R \to
\infty $ and the string pressure $ p_d \simeq E/2 \to + \infty $ ,
is {\bf positive}.
\item (iii) {\bf Stable} for $ R \to \infty $, (corresponding to $\eta
\to \infty , \tau\to \infty, S = $ constant ), for which the string
energy is $ E_s = $ constant and the string pressure vanishes $ p_s = 0 $ .
\end{itemize}

Here the indices $(u,d,s)$ stand for `unstable', `dual' and `stable'
respectively. The behaviours (i) and (ii) are related by the duality
transformation  $ R \leftrightarrow 1/R $ , the case (ii) being
invariant under duality. In the three cases, we find perfect fluid relations
(\ref{flup}) with the values of $\gamma$ :
\begin{equation}
\gamma_u = 1/2 \quad , \quad \gamma_d = 3/2 \quad , \quad\gamma_s = 1~.
\label{glup}
\end{equation}
For a perfect gas of strings on a comoving volume $ R^2 $, the energy
density $ \rho $ is proportional to $ E / R^2 $, which yields the
scaling  $ \rho_u = u \, R^{-1} , \rho_d = d \, R^{-3} , \rho_s = s\, R^{-2}$.
All densities and pressures obey the continuity equation (\ref{contA})
as it must be.

The $ 1+1 $ and $ 2+1 $ string solutions here described exist in any
spacetime dimension. Embedded in D-dimensional universes, the
 $ 1+1 $ and $ 2+1 $  solutions  describe straight strings and
circular strings, respectively. In D-dimensional spacetime,
 strings may  spread in $ D - 1 $ spatial dimensions. Their
treatment has been done asymptotically in ref.\cite{gsv}. We have
three general asymptotic behaviours:
\begin{itemize}
\item (i) {\bf unstable}  for $ R \to \infty $
with
$ \rho_u = u \, R^{2-D} ,\;   p_u =  -\rho_u/(D-1) < 0 $
\item (ii) {\bf Dual} to unstable  for $ R \to 0 $
with  $ \rho_d = d \, R^{-D},\; p_d = \rho_d/(D-1) > 0  $ .
\item (iii) {\bf Stable} for $ R \to \infty $,
with $\rho_s = s \, R^{1-D} , \;  p_s = 0 $ .
\end{itemize}

We find perfect fluid relations with the factors
\begin{equation}
\gamma_u = {{D-2}\over {D-1}} \quad , \quad \gamma_d =
  {D\over {D-1}} \quad , \quad\gamma_s = 1~.
\label{glud}
\end{equation}
This reproduces  the two dimensional and three dimensional results
for $ D = 2 $ and $ D = 3 $, respectively. The stable regime is absent
for $ D = 2 $ due to the lack of string transverse modes there.

The dual strings behave as {\it radiation} (massless particles) and the
stable strings are similar to {\it cold matter}. The unstable strings
correspond to the critical case of the so called {\it coasting universe}
\cite{ell,tur}. That is, classical strings provide a {\it concrete}
realization of such cosmological models.
The `stable' behaviour is called  'string stretching'
in the cosmic string literature \cite{twbk,vil}.

Strings continuously evolve from one type of behaviour to another, as
is explicitly shown by our solutions \cite{prd,dls}. For
intermediate values of $ R $, the string equation of state is clearly
more complicated but a formula of the type:
\begin{equation}
\rho = \left( u \; R + {{d} \over R} + s \right) {1 \over
{R^{D-1}}} ~~~,~~~
p  = {1 \over {D-1}} \left( {d \over R} -  u \; R\right) {1 \over
{R^{D-1}}} \label{rope}
\end{equation}
 is qualitatively
correct for all $ R $ and becomes exact for $ R \to 0 $ and $ R \to
\infty $ . We stress here that we obtained the string equation of
state from the exact string evolution in cosmological spacetimes.
Inserting the equation of state (\ref{rope}) in the Einstein-Friedmann
equations of general relativity, we obtain a self-consistent solution
for $ R $ as a monotonically increasing function of the cosmic time $
X^0 $ [eq.(\ref{inte})]. This string dominated universe starts at $
X^0 = 0 $ with a radiation dominated regime $
R(X^0) \buildrel{X^0 \to 0}\over \simeq C_D \, (X^0)^{2 \over D}$,
then the universe expands for large $ X^0 $ as
$ R(X^0) \buildrel{X^0 \to \infty}\over \simeq C'_D ~
 (X^0)^{2 \over {D-2}}$, which is faster than (cold) matter
dominated universes (where $ R \sim   (X^0)^{2 \over {D-1}}$).
For example, at $ D = 4 ,\;  R $ grows linearly with $ X^0 $, while for
matter dominated universes   $ \, R \sim   (X^0)^{2 \over 3}$.

The unstable string solutions are called in this way since their
energy and invariant length grow as $R$ for large $R$. However, it
must be clear that as {\it classical} string solutions they {\bf never
decay}.
At the quantum level  {\bf strings decay} through splitting and
we take these processes  into account. Unstable strings
split  with a probability proportional to $R$ as $R$ increases.
Long strings often split into massless strings
(radiation) and a long piece \cite{mtw};
that is, into a `dual' string  with energy
$\sim \, 1/R$ and a long string with energy $\sim \, R \,$ for large R.
Now $u$ and $d$ in eqs.(\ref{rope}) become functions of the time
$X^0$ which we find in eqs.(\ref{dent}). String decay {\bf changes
drastically} the large $R$ behaviour since the unstable string density
$u(X^0)$ decreases exponentially. Taken into account string decay,
our self-consistent solution $R(X^0)$ yields the realistic matter
behaviour $R(X^0)  \buildrel{X^0 \to \infty}\over \sim (X^0)^{2 \over
{D-1}}$ . Through quantum string splitting, the unstable strings do
not dominate anymore for large $R$. The {\it stable} strings (which
behave as cold matter) are those surviving for $R \to \infty$ . The
`dual' strings are not significantly affected by string decay and
give again
$R(X^0) \buildrel{X^0 \to 0}\over \simeq C_D \, (X^0)^{2 \over D}$,
 the radiation type behaviour.
For intermediate $R$, the three types of string behaviours (unstable,
dual and stable) are present. Their cosmological implications as well
as those associated with string decay deserve investigation. For a
thermodynamical gas of strings the temperature $T$ as a function of
$R$, scales as $1/R$ for small $R$ (the usual radiation behaviour).
Without string decay, $T$ {\bf grows} as $R$ for large $R$
[eq.(\ref{tempe})]. This strange behaviour disapears when string
splitting is taken into account as shown by eq.(\ref{temcs}).

For the sake of completeness we analyze the effective string
equations. These equations have been extensively treated in the
literature \cite{eqef} and they are not our central aim.

It must be noticed that there is no satisfactory
derivation of inflation in the context of the effective string equations.
 This does not mean that string
theory is not compatible with inflation, but that the effective string
action approach {\it is not enough} to describe inflation. The
effective string equations are a low energy field theory approximation
to string theory containing only the {\it massless} string modes.
The vacuum energy scales to start inflation are typically of the order
of the Planck mass where the effective string action approximation
breaks down. One must also consider the {\it massive} string modes (which
are absent from  the effective string action) in order to properly get
the cosmological condensate yielding inflation.
De Sitter inflation does not emerge as a solution of the
the effective string equations.

We find that a de Sitter spacetime with Lorentzian signature
self-sustained by the strings necessarily requires a constant imaginary
part $ \pm i \pi $ in the dilaton field $ \Psi $ , telling us that the
gravitational constant $ G \sim  e^{ \Psi } < 0 $ , here describes {\it
antigravity} [see eqs.(\ref{anti})]. In the euclidean signature case
$ (+++ \ldots +) $, we find a constant curvature  solution
self-sustained by the strings with
a real dilaton and   $ G \sim  e^{ \Psi } > 0 $, but of anti-de Sitter
type. Both  self-consistent solutions, de Sitter and anti-de Sitter
are mapped one into another by the transformation (\ref{traf}).

The outline of this paper is as follows. Section II deals with the
string propagation and the string energy-momentum tensor in
cosmological spacetimes. (In sections II.A, II.B and II.C we treat the
$1+1,~2+1$ and $D$-dimensional cases respectively and derive the
corresponding string equations of state). In section III we treat
self-consistent string cosmology including the string equations
of state and the quantum string decay. (Section III.A deals with
general relativity, III.B with the quantum string decay and III.C with
the effective (beta functions) string equations).

\section{\bf String propagation in cosmological spacetimes}

We obtain  in this section  physical string properties
from the exact string solutions in cosmological spacetimes.

We consider strings in spatially homogeneous and isotropic
universes with metric

\begin{equation}
ds^2 = (dX^0)^2 - R(X^0)^2 \sum_{i=1}^{D-1}(dX^i)^2
\label{met}
\end{equation}

where $X^0$ is the cosmic time. In terms of the conformal time

\begin{equation}
\eta = \int^{X^0} \frac{d X^0}{R(X^0)}~,
\label{tco}
\end{equation}

the metric (\ref{met}) takes the form

\begin{equation}
ds^2 = R(\eta)^2 \left[ (d\eta)^2 - \sum_{i=1}^{D-1}(dX^i)^2 \right]
\label{metc}
\end{equation}

The classical string equations of motion can be written  here as
\begin{eqnarray}
\partial^2X^0& - R(X^0)\; {{{\rm d}R}\over{\,{\rm d}X^0}}\;
\sum_{i=1}^{D-1}(\partial_\mu X^i)^2=0\,,\label{eqmov}\\
\partial_\mu&\left[R^2 \partial^\mu X^i\right]=0\,,
\qquad 1\leq i\leq D-1\,,\nonumber
\end{eqnarray}
 and the constraints are

\begin{equation}
T_{\pm\pm}=(\partial_\pm X^0)^2 - R(X^0)^2 \; (\partial_\pm X^i)^2=0\,.
\label{vincu}
\end{equation}

The spacetime string energy-momentum tensor is given by:

\begin{eqnarray}
\sqrt{-G}~ T^{AB}(X) = \frac{1}{2\pi \alpha'} \int d\sigma d\tau
\left( {\dot X}^A {\dot X}^B -X'^A X'^B \right)
\delta^{(D)}(X - X(\sigma, \tau) )
\label{tens}
\end{eqnarray}
where dot and prime stands for $\partial/\partial\tau$ and
 $\partial/\partial\sigma $, respectively.

As we will see below, $T^{AB}(X)$ takes the fluid form for string
solutions allowing us to define the string pressure $p$ and energy
density $\rho$ :

\begin{eqnarray}
T_A^B = \left(\begin{array}{rrrr}
\rho & 0 & \cdots & 0 \\
0  & -p & \cdots & 0 \\
\cdots & \cdots  & \cdots & 0 \\
0  & 0  &  \cdots & -p \end{array}\right)
\label{tflu}
\end{eqnarray}

Notice that the continuity equation

\begin{eqnarray}
D^A\;T_A^B = 0 \nonumber
\end{eqnarray}

takes here the form

\begin{equation}
\dot{\rho} + (D-1)\, H\, (p + \rho ) = 0
\label{cont}
\end{equation}

where $ H \equiv {{1}\over {R}} {{dR}\over{dX^0}} $.

For an equation of state of the type of a perfect fluid, that is

\begin{equation}
p = (\gamma - 1 )\;\rho \qquad , \qquad \gamma = {\rm~constant},
\label{eflu}
\end{equation}

 eqs.(\ref{cont}) and (\ref{eflu}) can be easily integrated with the
result

\begin{equation}
\rho = \rho_0~R^{\,\gamma(1-D)}~~.
\label{ror}
\end{equation}

For $\gamma = 1$ this corresponds to cold matter $(p = 0)$ and
for $\gamma = \frac{D}{D-1} $ this describes radiation with
$p = {{ \rho}\over{D-1}}$.

\subsection{1+1 Dimensional Universes}

Let us start by considering strings in this simpler case. Here $ D =
2$ and the metric (\ref{met}) takes the form

\begin{eqnarray}
ds^2 = (dX^0)^2 - R(X^0)^2 \;(dX)^2
\nonumber
\end{eqnarray}

It is convenient to start by solving the constraints (\ref{vincu})

\begin{equation}
(\partial_\pm X^0)^2 = R(X^0)^2 \; (\partial_\pm X)^2 \, .
\label{vincud}
\end{equation}

They reduce to
\begin{equation}
\partial_\pm X^0 = \epsilon_{\pm}~~ R(X^0) \; \partial_\pm X \, .
\label{raiz}
\end{equation}

where $ \epsilon_{\pm}^2 = 1$ . Using the conformal time (\ref{tco})
, eq.(\ref{raiz}) yields

\begin{eqnarray}
\partial_\pm ( \eta -  \epsilon_{\pm} X ) = 0 \, .
\nonumber
\end{eqnarray}

We find a first family of solutions choosing
 $ \epsilon_{\pm} = \pm 1 $ . Then
\begin{equation}
\eta + X = \phi(\sigma + \tau) ~~,~~ \eta - X = \chi(\sigma - \tau)
\label{sold}
\end{equation}
Where $\phi$ and $\chi$ are arbitrary functions of one variable.
It is easy now to check that eq.(\ref{sold}) fulfills the string
equations of motion (\ref{eqmov}).

The solution (\ref{sold}) is analyzed in detail for de Sitter
spacetime $( R(X^0) = e^{H X^0} )$ in ref.\cite{prd} where the
global topology of the space is taken into account.

Since one can always perform conformal transformations
\begin{eqnarray}
\sigma + \tau \to f(\sigma + \tau) \quad , \quad
\sigma - \tau \to g(\sigma - \tau)\quad , \nonumber
\end{eqnarray}

with arbitrary functions $f$ and $g$ , the solution (\ref{sold})
has no degrees of freedom other than topological ones.

Let us compute the energy momentum tensor for the string
solution  (\ref{sold}). We find from eqs.(\ref{tens}) and  (\ref{sold}),

\begin{eqnarray}
\sqrt{-G}~ T^{00}(X) & = & \frac{1}{2\pi \alpha'} \int d\sigma d\tau
\left( {\dot \eta}^2 -{\eta'}^2 \right)
\delta(\eta - \eta(\sigma, \tau) )\delta(X - X(\sigma, \tau) ) \nonumber\\
& = & \frac{1}{2\pi \alpha'} {{ {\dot \eta}^2 -{\eta'}^2}\over J}
\end{eqnarray}

where $ J = \frac{\partial(X,\eta)}{\partial(\sigma,\tau)} $ is the
jacobian. From eq.(\ref{sold}) we find $J = -\chi'\,\phi'$ and $
 {\dot \eta}^2 -{\eta'}^2  =  -\chi'\,\phi'$. Then,
\begin{eqnarray}
\sqrt{-G}~ T^{00}(X) = \frac{1}{2\pi \alpha'} ~.
\nonumber
\end{eqnarray}
 We analogously find $ {\dot X}^2 -{X'}^2  =  -\chi'\,\phi'$. Then
\begin{eqnarray}
\sqrt{-G}~ T^{11}(X) = -\frac{1}{2\pi \alpha'} ~~,~~ T^{01}(X)=0~.
\nonumber
\end{eqnarray}

That implies,
\begin{equation}
\rho = \frac{1}{2\pi \alpha'} ~~,~~ p= -\frac{1}{2\pi \alpha'}~~,~~
p + \rho = 0 .
\label{rop}
\end{equation}

We find a constant energy density and a constant {\bf negative}
pressure.
They exactly fulfill the continuity equation (\ref{cont}).
These results hold for {\bf arbitrary} cosmological spacetimes
in $1 + 1$ dimensions. That is, for  arbitrary factors $R(X^0)$.
In particular they are valid for strings winded $n$-times around the de Sitter
universe \cite{prd}.

A second family of string solutions follows from eq.(\ref{vincud})
by choosing
\begin{equation}
\eta = \pm X + C_{\pm}
\label{etx}
\end{equation}
where $C_{\pm}$ is a constant.
Then, the string equations of motion (\ref{eqmov})
become
\begin{equation}
\partial_{\mu}\left[R(\eta)^2\;\partial^{\mu} \eta \right]=0
\label{buf}
\end{equation}
Using eq.(\ref{etx}), we find that the energy-momentum tensor
(\ref{tens}) is traceless for this string solution:
\begin{equation}
T^{00} = T^{11} \quad,\quad {\rm that~is~~} p = \rho
\label{duud}
\end{equation}
Let us call
\begin{eqnarray}
V(\eta) = \int^{\eta}R^2(x) dx \nonumber
\end{eqnarray}
Then, eq.(\ref{buf}) implies that
\begin{eqnarray}
\left( {{\partial^2}\over{ \partial \tau ^2}}-
{{\partial^2}\over{ \partial \sigma ^2}} \right) V(\eta) = 0 . \nonumber
\end{eqnarray}
The general solution $\eta = \eta(\sigma, \tau) $ is implicitly
defined by:
\begin{eqnarray}
\int^{\eta}R^2(x) dx = A(\sigma -  \tau) + B(\sigma + \tau)~.\nonumber
\end{eqnarray}
where $A(x)$ and $B(x)$ are arbitrary functions. Upon a conformal
transformation, without loss of generality we can set
\begin{eqnarray}
A(\sigma -  \tau) + B(\sigma + \tau) ~~\Rightarrow ~~\tau~. \nonumber
\end{eqnarray}
Hence, $\eta$ is in general a function solely of
$\tau$ with
\begin{eqnarray}
\tau = \int^{\eta}R^2(x) dx ~~ {\rm or} ~~{{d\eta} \over {d\tau}} =
{1 \over {R^2(\eta)}}~.\nonumber
\end{eqnarray}
The energy-momentum tensor (\ref{tens}) takes for this solution the
form:
\begin{eqnarray}
T_A^B = {1 \over {\alpha'R^2}}
\delta(\eta \mp X - C_{\pm})\left(\begin{array}{rr}
1 & 0 \\
0  & -1 \end{array}\right)
\nonumber
\end{eqnarray}
The $\delta(\eta \mp X - C_{\pm})$ characterizes a localized object
propagating on the characteristics at the speed of light.
This solution describes a massless point particle since it has been
possible to gauge out the $\sigma$ dependence .

In summary, the two-dimensional string solutions in cosmological
spacetimes generically obey perfect fluid equations of state with
either
\begin{equation}
p = -\rho ~~(\gamma = 0) \qquad {\rm or}
\qquad p = + \rho ~~(\gamma =2)
\label{prou}
\end{equation}
The respectives energy densities being
\begin{equation}
\rho = \rho_u ~~(\rho_u ={\rm constant)~for~}\gamma = 0\quad {\rm or} \quad
\rho = {{u}\over {R^2}} ~~(u={\rm constant)~for~}\gamma = 2 .
\label{rodu}
\end{equation}
These behaviors fulfill the continuity equation (\ref{cont}) for  $D= 2$ .

\subsection{2+1 Dimensional Universes}

In $2+1$ dimensions, a large class of exact solutions describing
one string and multistrings
has been found in de Sitter universe
\cite{dms} -\cite{igor}.
For power-like expansion factors $ R(\eta)^2 = A \eta^k ~,~
(k \neq -2) $ only ring solutions are known\cite{din}.
($k = -2 $ corresponds to de Sitter spacetime).
Ring solutions correspond to the Ansatz \cite{din}:
\begin{eqnarray}
X^0 & = & X^0(\tau) \nonumber \\
X^1 & = & f(\tau) \cos \sigma \\ \nonumber
X^2 & = & f(\tau) \sin \sigma \label{ani}  \nonumber
\end{eqnarray}
The total energy of one string is then given by (recall $G^{00}=1$)
 \begin{eqnarray}
E(X^0) = \int d^{D-1}X \sqrt{-G}~T^{00}(X) = {1 \over {\alpha'}}
{{dX^0} \over {d \tau}}\nonumber
\end{eqnarray}
More generally, the energy-momentum tensor integrated on a volume
that completely encloses the string, takes the form \cite{ijm}
\begin{eqnarray}
\Theta^{AB}(X) & = & \frac{1}{2\pi \alpha'} \int d\sigma d\tau
\left( {\dot X}^A {\dot X}^B -X'^A X'^B \right)
\delta(X^0 - X^0(\tau) ) \nonumber \\
& = & \frac{1}{2\pi \alpha' |{\dot X}^0(\tau)|}
\int_0^{2\pi}d\sigma \left[ {\dot X}^A {\dot X}^B -X'^A X'^B \right]_
{\tau = \tau(X^0)}  \nonumber
\end{eqnarray}
For multistring solutions, one must sum over the different roots
$\tau_i$ of the equation $ X^0 = X^0(\tau) $ , for a given $X^0$.

We find for the ring ansatz eq.(\ref{ani}):
\begin{eqnarray}
\Theta^{00}(X) & = & E(X^0) \nonumber \\
\Theta^{11}(X) & = & \Theta^{22}(X) =
{1 \over {2 \alpha' |{\dot X}^0(\tau)|}}
[{\dot f}^2 - f^2 ] \nonumber \\\nonumber
\Theta^{01}(X) & = & \Theta^{02}(X) = \Theta^{12}(X) = 0 ~.
\label{tent}\\ \nonumber
\end{eqnarray}
That is,

\begin{eqnarray}
\Theta_A^B = \left(\begin{array}{rrr}
E & 0  & 0 \\
0  & -P  & 0 \\
0  & 0   & -P \end{array}\right)
\nonumber
\end{eqnarray}

where,
\begin{equation}
E = {1 \over {\alpha'}}~{\dot X}^0(\tau) \quad,\quad
P = {{R(\tau)^2} \over {2 \alpha' |{\dot X}^0(\tau)|}}~~
[{\dot f}^2 - f^2 ]
\label{epa}
\end{equation}

Let us first consider ring strings in de Sitter universe
\cite{dms} - \cite{dls}. There are three different
asymptotic behaviors: stable, unstable and its dual.
The unstable regime appears for $ \eta \simeq {{\tau}\over H}
\to 0 $.
{}From eqs.(\ref{epa}) and ref.\cite{dms}, we find
\begin{eqnarray}
E(\tau) &\buildrel{\tau \to 0^-}\over=& {1\over {\alpha'H }}
\left( {1 \over {|\tau|}} + 1 \right) \simeq {1\over {\alpha'H }}
\left( R(\tau) + 1 \right) \to +\infty \nonumber \\
P(\tau) &\buildrel{\tau \to 0^-}\over= & -{1\over {2 \alpha 'H |\tau|}}
\simeq -{{R(\tau)}\over {2 \alpha'H }}\to -\infty \label{desu}
 \nonumber
\end{eqnarray}
(Here $H$ stands for the Hubble constant).
The invariant string size grows as $R/H$ in this unstable regime.
Notice that the pressure is {\bf negative} for the unstable strings
and proportional to the expansion factor $R$.  In this
regime we also see that
\begin{equation}
P   \buildrel{R \to \infty}\over= -{E \over 2} +
{1 \over {2 \alpha' H}} + O(1/R)\to -\infty
\label{peun}
\end{equation}
with $E =  {1\over {\alpha'H }} \left( R(\tau) + 1 \right) $.
The regime dual to the unstable regime appears when the conformal time $\eta$
tends to infinity. For the solution  $q_-(\sigma,\tau)$ of ref.\cite{dms}
in de Sitter universe, $\eta$ diverges for finite $\tau \to \tau_0$ as
\begin{eqnarray}
\eta   \buildrel{\tau \to \tau_0}\over= {{6\;e^{-\tau_0} \over
{\tau -\tau_0}}} + O(1) \to +\infty \nonumber
\end{eqnarray}
Here
\begin{eqnarray}
\tanh{{\tau_0}\over{\sqrt{2}}} = {1 \over {\sqrt{2}}} \quad,\quad
\tau_0 = 1.246450.... \nonumber
\end{eqnarray}
Then,
\begin{eqnarray}
R(\tau) & \buildrel{\tau \to \tau_0}\over= & {{e^{\tau_0}} \over 6}\,
(\tau -\tau_0) \to 0^+ \nonumber \\
E(\tau) &\buildrel{\tau \to \tau_0}\over=& {1\over {\alpha'H (\tau -\tau_0)}}
= { 0.5796... \over {\alpha'H R}} \to +\infty \nonumber \\
P(\tau) &\buildrel{\tau \to \tau_0}\over=& {1\over {2\alpha'H (\tau -\tau_0)}}
= E/2 \to +\infty \label{dual}   \nonumber
\end{eqnarray}
We call dual to this regime since it appears related to the unstable regime
(\ref{desu}) through the exchange $ R \leftrightarrow 1/R $ .
The invariant string size tends to $ {1\over  H }$ in this regime.

In the stable regime, $\tau \to \infty$,
(and the cosmic time $ X^0 \simeq {{\tau}\over H} \to \infty $),
from eq.(\ref{epa}) and ref.\cite{dms}, we find
\begin{equation}
E(\tau) \buildrel{\tau \to \infty}\over= {1\over {\alpha'H }}
\quad , \quad
P(\tau) \buildrel{\tau \to \infty}\over= \frac{1+\sqrt{2}}{2 H}
e^{-\tau\sqrt{2}} \to 0 .
\label{esta}
\end{equation}

For stable strings in de Sitter universe, the pressure
 is positive and vanishes asymptotically, and the invariant string size
tends to $ {1\over {\sqrt{2} H }}$.

The solution  $q_-(\sigma,\tau)$ in ref.\cite{dms} describes
two ring strings: A stable string for $ q_0 \to +\infty
( \tau \to \infty ) $
($q_0$ being the hyperboloid time-like coordinate) and a unstable one for
 $q_0 \to -\infty  ( \tau \to \tau_0 )$.
 The pressure $P$ depends on $\tau$ ; it is
negative for $ q_0 < l$ and positive for $q_0 > l$, where
$l = -1.385145...$.

It can be noticed that the behaviour eq.(\ref{desu}) for the energy can
be interpreted as an unstable piece $ R/(\alpha' H) $ {\it plus} a
stable one $ 1/(\alpha' H) $ . The constant term   $ 1/(\alpha' H) $
is precisely  the energy for the stable solution eq.(\ref{esta}).

\bigskip

Let us now study the energy and pressure for the ring string solutions
in FRW and power-type inflationary universes considered in
\cite{din}. In terms of the conformal time $\eta$ , we have as expansion
factor $ R^2(\eta) = A \, \eta^k $.

Near $\eta = 0$ , two types of behavior were found \cite{din}.
The first one is a linear behavior
\begin{eqnarray}
\eta  & \buildrel{\tau \to \tau_0}\over= & \tau -\tau_0 +
O(\tau-\tau_0)^2 \nonumber \\\label{etan}
f(\tau) & \buildrel{\tau \to \tau_0}\over= &
1 - {{(\tau-\tau_0)^2}\over{2(k+1)}} ~~{\rm for~}k>0{\rm~
and~} k < -1  \\
{\rm and~} f(\tau) & \buildrel{\tau \to \tau_0}\over= &
1 + c ~(\tau-\tau_0)^{1-k}  ~~{\rm for~} -1<k<0.\nonumber
\end{eqnarray}
Eqs.(\ref{etan}) describe a collapsing (expanding) string for $ k > 0~
 ( k < 0 ) $ with invariant size $ \simeq (\tau-\tau_0)^{k/2} $.
That is, in standard FRW universes $(k>0)$, the string size goes to
zero for $\eta \simeq (\tau-\tau_0) \to 0 $ as the universe radius
$R \simeq (\tau-\tau_0)^{k/2} \to 0 $. In inflationary universes
$(k < 0)$ , the string size grows indefinitely for
 $\eta \simeq (\tau-\tau_0) \to 0 $ as the universe radius
$R \simeq (\tau-\tau_0)^{k/2} \to +\infty $. The growing of the proper
string size for $\eta  \simeq \tau -\tau_0 \to 0 $ is a typical
feature of string instability.

Using eqs.(\ref{epa}) and (\ref{etan}), the energy and pressure take the
form,
\begin{eqnarray}
E(\tau) & \buildrel{\tau \to \tau_0}\over= &
{{\sqrt{A}}\over  {\alpha'}}~(\tau-\tau_0)^{k/2}= R/\alpha'
\quad , \nonumber \\
P(\tau) & \buildrel{\tau \to \tau_0}\over= &
-{{\sqrt{A}}\over  {2\alpha'}}~(\tau-\tau_0)^{k/2}= -R/(2 \alpha')
\quad . \label{epun}
\end{eqnarray}
That yields
\begin{equation}
P \buildrel{\tau \to \tau_0}\over= - E/2
\label{eqes}
\end{equation}
Notice that eq.(\ref{eqes}) holds both for $R\to\infty$ in power-like
inflationary universes ($k<0$) and for $R\to 0$ in
standard FRW universes
($k>0$).

Eq.(\ref{eqes}) is also valid  in de Sitter universe for unstable
strings [eq.(\ref{peun})].
In all these cases strings exhibit {\bf negative} pressure.

The second behavior present near $\eta = 0$ is \cite{din}
\begin{equation}
\eta   \buildrel{\tau \to \tau_0}\over=  (\tau -\tau_0)^{1/(k+1)}
\quad , \quad
f(\tau) \buildrel{\tau \to \tau_0}\over= f_0 \pm  (\tau -\tau_0)^{1/(k+1)}
\label{eand}
\end{equation}
where $f_0$ must be set equal to zero for $-1 < k < 0 $.
The invariant string size behaves for $ \tau \to \tau_0 $ as,
\begin{eqnarray}
S(\tau) & \buildrel{\tau \to \tau_0}\over=  &
\sqrt A~(\tau -\tau_0)^{{k+2}\over {2(k+1)}}  \quad {\rm for~} k< 0
\nonumber \\ \nonumber
S(\tau) & \buildrel{\tau \to \tau_0}\over=  &
\sqrt A~(\tau -\tau_0)^{{k}\over {2(k+1)}}  \quad {\rm for~} k> 0
\nonumber
\end{eqnarray}
This solution describes a string that collapses for $k > 0$
and for $k < -2$ , that is, both for standard FRW and inflationary cosmologies.
Here, the expansion factor tends to zero as
\begin{eqnarray}
R(\tau) \buildrel{\tau \to \tau_0}\over=
\sqrt{A}~(\tau -\tau_0)^{{k}\over {2(k+1)}} \to 0.\nonumber
\end{eqnarray}
when $k > 0$ and when $k < -1$ .

That is, in this case, for small radius $ R \to 0 $ of both standard
FRW and inflationary cosmologies, strings collapse.

{F}rom eq.(\ref{epa}) we find that
\begin{eqnarray}
P = {A \over {2(k+1)\alpha' R}}\to +\infty \quad , \quad
E = {A \over {(k+1)\alpha' R}} \to +\infty \nonumber
\end{eqnarray}
That is,
\begin{equation}
P \buildrel{\tau \to \tau_0}\over= E/2
\label{duat}
\end{equation}
Notice that the pressure is here {\bf positive}.
This second behaviour is related by duality ($ R \leftrightarrow 1/R $)
to the first  behaviour described by eqs.(\ref{epun})-(\ref{eqes}).

For large $\tau$ , the ring strings exhibit a stable
behaviour\cite{din},
\begin{eqnarray}
\eta(\tau)&\buildrel{{\tau\to+\infty}}\over=
&\tau^{2/(k+2)}\,,\nonumber \\
X^0(\tau)&\buildrel{{\tau\to+\infty}}\over= &
{{2\sqrt{A}}\over{k+2}}\;\tau ~, \nonumber \\
f(\tau)&\buildrel{{\tau\to+\infty}}\over= &{2\over{k+2}} \; \tau^{-k/(k+2)}
\cos(\tau+\varphi)\,,\label{solasim}\\
\nonumber
\end{eqnarray}
where $ \varphi $ is a constant phase and the oscillation amplitude has been
 normalized. For large $\tau $ , the energy and pressure of the
solution (\ref{solasim}) behave as
\begin{eqnarray}
E(\tau) & \buildrel{{\tau\to+\infty}}\over= & {{2\sqrt{A} } \over {\alpha'
(k+2)}} =  {\rm constant~} \quad, \nonumber \\
P (\tau) & \buildrel{{\tau\to+\infty}}\over= & -{{\sqrt{A} } \over {\alpha'
(k+2)}} \cos(2\tau + 2\varphi) \to 0 \quad .
\label{epes}
\end{eqnarray}

This is the analog of the stable behaviour (\ref{esta}) in de Sitter
universe for $ \tau \to \infty $.
Notice that eqs.(\ref{epes}) hold both for standard FRW
$( k > 0 )$ and inflationary spacetimes $(k < 0)$ .

For power-like inflationary universes with $ k < -1 $ a special exact
ring solution exist \cite{din} with
\begin{eqnarray}
\eta = C \, \exp{{{\tau} \over {\sqrt{-k-1}}}} \quad , \quad
f(\tau) = { C \over {\sqrt{-k}}} \exp{{{\tau} \over {\sqrt{-k-1}}}}
\label{exot}
\end{eqnarray}
where $ C $ is an arbitrary constant. For $ k = - 2 $ (de Sitter
universe) this is the solution $q^{(o)}(\sigma, \tau)$ in
ref.\cite{dms} which has constant string size.

For this solution we find
\begin{eqnarray}
E & = & \frac{1}{\alpha'}\sqrt{{A \over {-k-1}}}\;
\exp\left[{{{\tau (k+2)} \over {2\sqrt{-k-1}}}}\right] = K \; R^{1 + 2/k}
\nonumber \\
P & = & -({1 \over 2}+{1 \over k})~E \label{eexo} \nonumber
\end{eqnarray}
where $ K $ is a constant.

This is a fluid-like equation of state with
$\gamma = {1 \over 2}-{1 \over k}$ . Notice that ${1 \over 2} < \gamma
< {3 \over 2} $. For this solution, the energy grows with $R$ as $R$ to
a power $ 1 + 2/k $ where, since $ k < - 1 $ ,
\begin{eqnarray}
- 1 < 1 + {2 \over k} < 1 . \nonumber
\end{eqnarray}
($E$ and $P$ of this solution are constants in de Sitter spacetime $[k
= -2$]).
This means that these special strings are subdominant both for $R \to \infty$
and for $ R \to 0 $ where the unstable strings ( $ E \simeq R $ )
and their dual  ($ E \simeq R^{-1} $)  dominate
respectively.

\bigskip

In summary, three asymptotic behaviors are exhibited by ring
solutions.

\bigskip

i) Unstable for $R \to \infty:~ E = {\rm (constant)~} R \to +\infty ~
,~ P = -E/2 \to -\infty,~S \simeq R \to \infty $.

\medskip

ii) Dual to unstable for $R \to 0:~ E = {\rm (constant)~} R^{-1} \to
+\infty ~,~ P = +E/2 \to +\infty , \\ ~
S \simeq R \to 0$ (except for de Sitter
where $S \to {1 \over H}$).

\medskip

iii) Stable for $R \to \infty:~ E = {\rm constant~} ,~ P = 0, ~
 S = {\rm constant~} $.

\bigskip

In addition we have the special behavior (\ref{eexo}) for $ k < -1 $.
Notice that the three behaviors i)-iii) appear for all expansion factors
$R(\eta)$ . The behaviours i) and ii) are related by the duality
transformation $ R \leftrightarrow 1/R $, the case iii) being invariant
under duality . In the three cases we find perfect fluid equations
[see eq.(\ref{eflu})] with different $ \gamma $ :
\begin{eqnarray}
\gamma_u = 1/2 ~~,~~ \gamma_d = 3/2 ~~,~~ \gamma_s = 1~~, \nonumber
\end{eqnarray}
where the indices $u, d$ and $s$  stand  for `unstable',
`dual' and `stable', respectively.
Assuming a perfect gas of strings on a volume $R^2$ , the energy
density $\rho$ will be proportional to $ E/R^2 $. This yields
the following scaling with the expansion factor using
the energies from i)-iii):
\begin{equation}
\rho_u = {\rm constant~} R^{-1} ~~,~~ \rho_d =  {\rm constant~} R^{-3} ~~,~~
\rho_s =  {\rm constant~} R^{-2} .
\label{rodd}
\end{equation}
All three densities and pressures obey the continuity equation
(\ref{cont}), as it must be.

The factor $1/2$ in the relation between $P$ and $E$ for the  cases i) and
ii) is purely geometric. Notice that this factor was one in $1+1$
dimensions [eq.(\ref{prou})].

\subsection{D-Dimensional Universes}

The solutions investigated in secs. II.A and II.B  can exist for any
dimensionality of the spacetime. Embedded in D-dimensional universes,
the $(1+1)$ solutions of section  II.A describe
straight strings, the $(2+1)$ solutions of section  II.B are circular rings.
 In D-dimensional spacetime, one expects string solutions spread in
$D-1$ spatial dimensions. Their treatment has been done
asymptotically in ref.\cite{gsv}. One finds for $\eta \simeq \tau -
\tau_0 \to 0$ ,  $ R \to \infty$,
\begin{equation}
P_u = -{1 \over {D-1}}\;\rho_u ~~~,~~~  \gamma_u = {{D-2}\over{D-1}}  ~~~
{\rm for~unstable~strings.}
\end{equation}
This relation coincides with eq.(\ref{rop}) for $D = 2$ and with
eqs.(\ref{peun}) and (\ref{eqes}) for $D = 3$.

The energy density scales with $R$ as
\begin{eqnarray}
\rho_u = u \; R^{2-D} ~~~,~~~ R \to \infty
\label{estin}
\end{eqnarray}
(where $u$ is a constant)
in accordance with eq.(\ref{rodu}) for $ D = 2 $ and with
eq.(\ref{rodd}) for $ D = 3 $.

For the dual regime, $ R \to 0$, we have:
\begin{equation}
P_d = +{1 \over {D-1}}\;\rho_d ~~,~~ \gamma_d = {D\over{D-1}} ~~
{\rm with~} \rho_d = d \; R^{-D}~~,~~ R \to 0 ~~,
\label{estdu}
\end{equation}
(where $d$ is a constant).
Eq.(\ref{estdu}) reduces to eq.(\ref{duud}) for $ D = 2 $  and to
eqs.(\ref{dual}) and (\ref{duat}) for $ D = 3 $. In this
dual regime strings have the same equation of state as massless radiation.

Finally, for the stable regime we have
\begin{equation}
P_s = 0 ~~,~~ \gamma_s = 1~~{\rm with~} \rho_s = s \; R^{1-D}
\end{equation}
(where $s$ is a constant).
This regime is absent in $ D = 2 $ and appears for  $ D = 3 $ solutions
in eqs.(\ref{esta}) and (\ref{epes}).
The lack of string transverse modes in $ D = 2 $ explains the absence
of the stable regime there.
The equation of state
for stable strings coincides with the one for cold matter.

In conclusion, an ideal gas of classical strings in cosmological universes
exhibit three different thermodynamical behaviours,
all of perfect fluid type:

\medskip

1) Unstable strings :
negative pressure gas with $\gamma_u = {{D-2}\over{D-1}} $

2) Dual behaviour : positive pressure gas similar to radiation ,
$\gamma_d = {D\over{D-1}} $

3) Stable strings : positive pressure gas similar to cold
matter, $ \gamma_s = 1$.

\medskip

The unstable string behaviour corresponds to the critical case of the
so-called coasting universe \cite{ell},\cite{tur}. In other words, classical
strings provide a {\it concrete} matter realization of such cosmological
model. Till now, no form of matter was known to describe coasting universes
\cite{ell}.

Finally, notice that strings continuously evolve from one type of
behaviour to the other two. This is explicitly seen from the string
solutions in refs.\cite{dms} - \cite{dls} . For example the string
described by $q_-(\sigma, \tau)$ for $ \tau > 0$ shows unstable
behaviour for $ \tau \to 0 $, dual behaviour for $ \tau \to \tau_0 =
1.246450...$ and stable behaviour for $ \tau \to \infty $ .

The equation of state for strings in four dimensional flat Minkowski
spacetime is discussed in ref.\cite{romi}. One finds the values
4/3, 2/3 and 1 for $\gamma$ by choosing  appropiate  values of the
average string velocity in chap. 7 of ref.\cite{romi}.

\section{\bf Self-consistent string cosmology}

In the previous section we investigated the propagation of test
strings in cosmological spacetimes. Let us now investigate how
the Einstein equations in General Relativity and
the effective equations of string theory (beta functions) can be verified
{\bf self-consistently} with our string solutions as sources.

We shall assume a gas of classical strings neglecting interactions as
string splitting and coalescing. We will look for cosmological
solutions described by metrics of the type (\ref{met}). It is natural
to assume that the background will have the same symmetry as the sources.
That is, we assume that the string gas is homogeneous, described by a
density energy $ \rho = \rho(X^0) $ and a pressure $ p = p(X^0) $.
In  the  effective equations of string theory we consider a space
independent dilaton field. Antisymmetric tensor fields wil be ignored.

\subsection{\bf String Dominated Universes in General Relativity
(no dilaton field)}

The Einstein equations for the geometry (\ref{met}) take the form

\begin{eqnarray}
{1 \over 2}~(D-1)(D-2)~ H^2 & = & \rho \quad, \nonumber \\
(D-2){\dot H} + p + \rho & = & 0 \quad .
\label{eins}
\end{eqnarray}

where $ H \equiv {{dR}\over{dX^0}}/ R $. We know $p$ and $\rho$ as
functions of $R$ in asymptotic cases. For large $ R $, the unstable
strings dominate [eq.(\ref{estin})] and we have
\begin{equation}
\rho  = u \; R^{2-D} ~~~,~~~p = -{{\rho} \over {D-1}}\;
\quad {\rm  for ~}R \to \infty
\end{equation}

For small $ R $, the dual regime dominates with

\begin{equation}
\rho  = d \; R^{-D} ~~~,~~~p = +{ {\rho}\over {D-1}}\;\quad {\rm
for ~} R\to 0
\end{equation}

We also know that stable solutions may be present with a contribution
$ \sim R^{1-D} $ to $ \rho $ and with zero pressure. For intermediate values
of $ R $ the form of $ \rho $ is clearly more complicated but a formula
of the type

\begin{equation}
\rho = \left( u \; R + {{d} \over R} + s \right) {1 \over
{R^{D-1}}} \label{rogen}
\end{equation}

with $u , d$ and $ s $ being positive constants, is qualitatively
correct for all $ R $ and becomes exact for $ R \to 0 $ and $ R \to
\infty $ .

The pressure associated to the energy density (\ref{rogen}) takes then
the form
\begin{equation}
p  = {1 \over {D-1}} \left( {d \over R} -  u \; R\right) {1 \over
{R^{D-1}}} \label{pgen}
\end{equation}

Inserting eq.(\ref{pgen}) into the Einstein-Friedmann equations
[eq.(\ref{eins})] we find
\begin{equation}
{1 \over 2}~(D-1)(D-2)~ \left({{d R}\over{dX^0}}\right)^2  =  \left( u \; R +
 {d \over R} + s \right) {1 \over
{R^{D-3}}} \label{enfri}
\end{equation}

We see that $R$ is a monotonic function of the cosmic time $X^0$.
Eq.(\ref{enfri}) yields

\begin{equation}
X^0 = \sqrt{{(D-1)(D-2)}\over 2}~
\int_0^R dR \; {{R^{D/2-1}} ~\over {\sqrt{ u \; R^2  + d + s\; R}}}
\label{inte}
\end{equation}

where we set $R(0) = 0$.

It is easy to derive the behavior of $R$ for $X^0 \to 0$ and for $X^0
\to \infty$.

For  $X^0 \to 0$, $ R  \to 0$, the term  $d/R$ dominates in
eq.(\ref{enfri}) and
\begin{equation}
R(X^0) ~ \buildrel{X^0 \to 0}\over \simeq ~ {D \over 2}
\left[{{2 d}\over {(D-1)(D-2)}}\right]^{1 \over D}~ (X^0)^{2 \over D}
\label{rcer}
\end{equation}
For $X^0 \to \infty$, $R\to \infty$ and  the term $u\,R$ dominates  in
eq.(\ref{enfri}). Hence,
\begin{equation}
R(X^0) ~ \buildrel{X^0 \to \infty}\over \simeq ~
\left[{{(D-2) u}\over {2(D-1)}}\right]^{1 \over {D-2}}~ (X^0)^{2 \over {D-2}}
\label{rinf}
\end{equation}

For intermediate values of $X^0~ ,~ R(X^0)$ is a continuous and
monotonically increasing function of $X^0$.

In summary, the universe starts at $ X^0 = 0 $ with a singularity of
the type dominated by radiation. (The string behaviour for $R \to 0 $
is like usual radiation ). Then, the universe expands monotonically
growing for large $ X^0 $ as $ R \simeq  [X^0]^{{2 \over {D-2}}} $ . This is
faster than (cold) matter dominated universes where
$ R \simeq  [X^0]^{{2 \over {D-1}}} $ . For example, for $ D = 4 $, $
R $ grows linearly with $ X^0 $ whereas for matter dominated universes
$ R \simeq  [X^0]^{2/3} $ .

As is clear already in eq.(\ref{enfri}) , for large $R$ , unstable
strings ($u\;R$) dominate over the stable strings ($s$). These stable
strings behave as cold matter. It must be noticed that the qualitative
form of the solution $R(X^0)$ does not depend on the particular
positive values of $ u , d $ and $ s $.

We want to stress that we achieve a {\bf self-consistent} solution of the
Einstein equations with string sources since the  behaviour
of the string pressure and density given by eqs.(\ref{rogen})-(\ref{pgen})
precisely holds in universes with power like $ R(X^0) $. Since we find
positive exponents $K$ in the solution  (\ref{rinf}) both for small and
large $R$, we can consistently ignore the solutions (\ref{exot}) which appear
for $ k < 0 $.
 It  can be   noticed that a linear
growing behaviour but for {\it all D } follows from the effective
string equations (including the dilaton)  without the
string sources \cite{tse}. Notice that for string dominated universes,
$ R \simeq  X^0 $ appears {\it only} in $ D = 4 $.
On the other hand, the behaviour  $ R \simeq  X^0 $ in $ D = 4 $ also
appears, but for  $  X^0 \to 0 $ \cite{fhh}, in the context of the
semiclassical Einstein equations using as  source the four dimensional
trace anomaly contribution to the energy-momentum tensor of
quantum matter (point particle) fields.

\subsection{\bf String Decay}

The unstable string solutions are called in this way since their
energy and invariant length grow as $R$ for large $R$. However, it
must be clear that as {\it classical} string solutions they {\bf never
decay}. The situation changes at the quantum level where strings may split and
coalesce. Quantum string splitting calculations in Minkowski spacetime
show that the splitting probability is proportional to the string length
\cite{pol,mtw}. This result holds at the critical dimension $(D = 26)$
but it should be true more generally in absence of conformal anomalies
(critical strings).

Since unstable strings are of invariant size proportional to $R$,
this suggests that they will split  with a growing probability as
$R$ increases. Through splitting, two or more shorter strings will be
produced which grow as $R$  if their size is
$\sim \, R $. However, long strings often split into massless strings
(radiation) and a long piece \cite{mtw};
that is, into a 'dual' string  with energy
$\sim \, 1/R$ and a long string with energy $\sim \, R$ for large R.
This type of process is important because it reduces  the number
$(u)$ of unstable strings per comoving volume.

We can  incorporate string decay in the string gas model of sec.3 A
by considering $u$ and $d$  as functions of time ($X^0$) in eqs.
(\ref{rogen})-(\ref{pgen}). From the above discussion the time
derivative of  $u$ must be
proportional to the unstable string length $R$ and to  $u$ itself:
\begin{equation}
{{du}\over{dX^0}} = - \, {\cal C}\; R(X^0) \; u(X^0)
\label{desc}
\end{equation}
where $ {\cal C} $ is a positive constant of the order of the string
coupling constant squared ($\sim e^{2\phi_0}$). The number $ ( d ) $ of dual
strings (massless quanta) per comoving volume
increases as $ R^{2} (-{{du}\over {dX^0}})$
since an unstable string with energy $R$ yields $R^2$ quanta with
energy $1/R$ each. That is,
\begin{equation}
{d\over{dX^0}} \,  d(X^0)=  {\cal C}\; R(X^0)^3 \; u(X^0)
\label{proq}
\end{equation}
Eqs.(\ref{desc})-(\ref{proq}) can be easily integrated:
\begin{eqnarray}
u(X^0) &= u(0) ~ \exp\left[-{\cal C}\int_0^{X^0}dt~R(t)\right] \cr
d(X^0) &= d(0) +  {\cal C}\; \int _0^{X^0}dt~R(t)^3 \; u(t) ~.
\label{dent}
\end{eqnarray}
 From eqs.(\ref{rogen}),(\ref{pgen}) and (\ref{dent}), we find for the total
string density and pressure the following expressions,
 \begin{eqnarray}
\rho =&  R(X^0)^{1-D}
\left\{ u(0) \; R(X^0) \; \exp{\left[-{\cal C}\int_0^{X^0}dt~R(t) \right]}
 +  {1 \over R(X^0)} \left[ d(0) +
  {\cal C}\; \int _0^{X^0}dt~R(t)^3 \; u(t) \right]
+ s \right\}  \nonumber \\ \cr
p  =& {1 \over {D-1}}~R(X^0)^{1-D}
\left\{ {1 \over R(X^0)}\left[d(0) +
  {\cal C}\; \int _0^{X^0}dt~R(t)^3 \; u(t)\right]
 -  u(0) \; R(X^0) \; \exp\left[-{\cal C}\int_0^{X^0}dt~R(t)\right]
 \right\} \label{pgend}
\end{eqnarray}
A dramatic change happens for large expansion factors $R(X^0)$ since
$u(X^0)$ decreases {\it exponentially}.

Notice that the pressure and energy density from eqs.(\ref{pgend}) obey
the continuity equation (\ref{cont}).

The Einstein-Friedmann equations [eqs.(\ref{enfri})] now become
\begin{eqnarray}
&{1 \over 2}~(D-1)(D-2)~ \left({{d R}\over{dX^0}}\right)^2  =
 \left\{ u(0) \; R(X^0) \; \exp\left[-{\cal C}\int_0^{X^0}dt~R(t)\right]
\right. \cr
&\left. +{1 \over R} \left[d(0) +
  {\cal C}\; \int _0^{X^0}dt~R(t)^3 \; u(t)\right]
+ s \right\} ~R(X^0)^{3-D}
\label{neinf}
\end{eqnarray}
We see that $R(X^0)$ is always a monotonic function of $X^0$. It is easy to
derive from eq.(\ref{neinf}) the behaviour of  $R(X^0)$ for small and for
large  $X^0$.

For  $X^0 \to \infty , R \to \infty $ and we find a matter dominated
regime:
\begin{equation}
R(X^0)  \buildrel{X^0 \to \infty}\over \simeq
\left[{{(D-1) s}\over {2(D-2)}}\right]^{1 \over {D-1}}~ (X^0)^{2 \over {D-1}}
\label{mdom}
\end{equation}
The unstable strings provide subdominant exponentially small
corrections:
\begin{equation}
O((X^0)^{2 \over {D-1}} ~ \exp[-A(X^0)^{{D+1}\over{D-1}}]) ~~
,~~{\rm with~}A = {\cal
C}~(s/2)^{{1\over{D-1}}} ~ \left({{D-1}\over{D-2}}\right)^{D\over{D-1}}.
\end{equation}

Taking into account the string decay, the stable strings (which behave
as cold matter) are those which remain for $ R \to \infty $, and yield
the realistic large $ R $ behaviour (\ref{mdom}).
The dual strings for $ R \to 0 $ are not substantially
affected by the string decay.
For $ X^0 \to 0 , R
\to 0 $ and we have again the radiation type behavior (\ref{rcer}).

For very small splitting rate $\cal{C}$, unstable strings will decay
very slowly and they can dominate $\rho$ for a while. Assuming this is
the case, we can neglect the stable and  dual pieces in
eq.(\ref{pgend}) and we find that $R(X^0)$ approaches a constant value
$R_1$. Namely,
\begin{equation}
R(X^0) = R_1 - A ~ e^{-{1 \over 2}{\cal C}\,X^0\, R_1 }~
,{\rm ~for}~X^0 \geq 1/[{\cal C}\,R_1] ~,~~u(0)\,R_1 \gg s \nonumber
\end{equation}
where $A$ and $R_1$ are constant depending on the initial conditions.
In other words, $R(X^0)$ reaches some plateau and then it grows again
following the matter dominated behaviour (\ref{mdom}).

\subsection{\bf Thermodynamics of strings in cosmological spacetimes}

Let us consider a comoving volume $ R^{D-1} $ filled by a gas of
strings. The entropy change for this system is given by:
\begin{equation}
T dS = d (\rho \; R^{D-1}) ~+ ~ p\;d(R^{D-1})
\label{dentr}
\end{equation}
The continuity equation (\ref{cont}) and (\ref{dentr}) implies that $ dS/dt $
 vanishes.  That is, the entropy per comoving volume stays constant in
time.
 Using now the thermodynamic relation \cite{wei}
\begin{equation}
{{dp}\over{dT}} = {{p+\rho}\over T}
\end{equation}
it follows \cite{romi} that
\begin{equation}
S = {{R^{D-1}}\over T}(p + \rho) + {\rm constant}
\label{entro}
\end{equation}

Let us first ignore the possibility of string decay. Then,
eq.(\ref{entro}) together with  eqs.(\ref{rogen}) and (\ref{pgen}) yields
the temperature as a function of the expansion factor $ R $. That is,
\begin{equation}
T = {1 \over S}\left\{ s + {1 \over {D-1}} \left[ {{D\; d}\over R} +
(D -2)\; u \; R \right] \right\}
\label{tempe}
\end{equation}
where $ S $ stands for the (constant) value of the entropy.

Eq.(\ref{tempe}) shows that for small $ R ,~ T $ scales as  $ 1/R $
whereas for large $ R $ it scales as $ R $. The small $ R $ behaviour
of $ T $ is the usual exhibited by radiation. On the contrary, for
large R the temperature {\bf grows} proportional to  R.
This strange behaviour is actually absent when string splitting is
considered.

Let us consider string decay. Inserting eq.(\ref{pgend})
into eq.(\ref{entro}) yields:
\begin{equation}
T = {1 \over S}\left\{ s + {1 \over {D-1}} \left[ {D\over R}
\left( d(0) +  {\cal C}\; \int _0^{X^0}dt~R(t)^3 \; u(t) \right)
+ (D - 2)\; u(0) \; R
\exp\left(-{\cal C}\int_0^{X^0}dt~R(t)\right) \right] \right\}
\label{temcs}
\end{equation}
For small $R,~ T$ scales here as $ 1/R $. For large $ R $, stable
strings ($s$) dominates and one must take into account the
temperature dependence of
$s$ \cite{romi} , in order to determine $T$ as a function of $R$.

\section{\bf Effective String  Equations with the String Sources Included}

Let us consider now the cosmological equations obtained from the low
energy string effective action including the string matter as a
classical source. In D spacetime dimensions, this action can be
written as
\begin{eqnarray}
S & = & S_1 + S_2 \cr
S_1 & = &  {1 \over 2} \int d^Dx \; \sqrt{-G} ~e^{-\Phi}~\left[\; R +
G_{AB}\; \partial^A\Phi~\partial^B\Phi~+ 2~U(G,\Phi) - c\;\right] \cr
S_2 & = &  -{1 \over {4\pi\alpha'}}\sum_{strings}\int d\sigma d\tau~
G_{AB}(X)~\partial_{\mu}X^A\;\partial^{\mu}X^B \qquad,
\label{accef}
\end{eqnarray}
Here $ A, B = 0, \ldots , D-1 $.
This action is written in the so called `Brans-Dicke frame' (BD)
or `string frame', in which matter couples to the metric tensor in the
standard way. The BD frame metric coincides with the sigma model
metric to which test strings are coupled.

Eq.(\ref{accef}) includes the dilaton field ($\Phi$) with a
potential $U(G,\Phi)$ depending on the dilaton and graviton
backgrounds; $ c $ stands for the central charge deficit or
cosmological constant term. The antisymmetric tensor field was not
included, in fact it is irrelevant   for the results obtained here.
Extremizing the action (\ref{accef}) with respect to $G_{AB}$ and $\Phi$
yields the equations of motion
\begin{eqnarray}
R_{AB} + \nabla_{AB}\Phi  + 2 \, {{\partial U}\over {\partial
G_{AB}}} - {{G_{AB}}\over 2} \left[ R + 2 \, \nabla^2\Phi -
(\nabla \Phi)^2 - c + 2 \, U \right] & = & e^{\phi}~T_{AB} \nonumber \\ \cr
R + 2 \, \nabla^2\Phi - (\nabla \Phi)^2 - c + 2 U -
{{\partial U}\over {\partial \Phi}} & = & 0 ~,
\label{eqef}
\end{eqnarray}
which can be more simply combined as
\begin{eqnarray}
R_{AB} + \nabla_{AB}\Phi  + 2 \; {{\partial U}\over {\partial
G_{AB}}} - G_{AB}\;
{\,{\partial U}\over {\partial \Phi}}  & = & e^{\Phi}~T_{AB}
\nonumber \\ \cr
R + 2 \; \nabla^2\Phi - (\nabla \Phi)^2 - c + 2 \, U -
{{\partial U}\over {\partial \Phi}} & = & 0
\label{eqeff}
\end{eqnarray}
Here $ T_{AB} $ stands for the energy momentum tensor of the strings
as defined by eq.(\ref{tens}).
It is also convenient to write these equations as
\begin{equation}
R_{AB}- {{G_{AB}}\over 2}\;R =   T_{AB} + \tau_{AB}
\end{equation}
where $ \tau_{AB} $ is the dilaton  energy momentum tensor :
\begin{eqnarray}
\tau_{AB} & = &  -\nabla_{AB}\Phi + {{G_{AB}}\over 2}
 \left[ 2 \, {{\partial U}\over {\partial \Phi}} - R \right] \nonumber
\end{eqnarray}
The Bianchi identity
\begin{eqnarray}
\nabla^A\left(R_{AB}- {{G_{AB}}\over 2}\;R \right)  & = & 0
\nonumber \cr
\end{eqnarray}
yields, as it must be, the conservation equation,
\begin{equation}
\nabla^A\left( T_{AB} + \tau_{AB}\right) = 0
\end{equation}
It must be noticed that eqs.(\ref{eqeff}) do not reduce to the
Einstein equations of General Relativity even when $ \Phi = U = 0 $.
Eqs. (\ref{eqeff}) yields in that case the
Einstein equations {\it plus} the condition $ R = 0 $.

\subsection{Effective String Equations in Cosmological Universes}

For the homogeneous isotropic spacetime
geometries described by eq.(\ref{met}) we  have
\begin{eqnarray}
R_0^0 & = & - (D-1) ( {\dot H} + H^2 ) \cr
R_i^k & = & - \delta_i^k ~ [ {\dot H} + (D-1)\,H^2 ] \cr
R  & = & - (D-1) ( 2\;{\dot H} + D\,H^2 ) .
\end{eqnarray}
where $H \equiv {1 \over R}\,{{dR}\over{dX^0}}$ .

The equations of motion (\ref{eqeff}) read
\begin{eqnarray}
{\ddot \Phi } - (D-1) ( {\dot H} + H^2 ) - {{\partial U}\over
{\partial \Phi}} & = &  e^{\Phi}~\rho  \nonumber \\ \cr
 {\dot H} + (D-1)\,H^2 - H \, {\dot \Phi }  + {{\partial U}\over
{\partial \Phi}} + { R \over {D-1}}{{\partial U}\over
{\partial R}} & = &  e^{\Phi}~p  \nonumber \\ \cr
2\;{\ddot \Phi } + 2(D-1)\,H \, {\dot \Phi }- {\dot \Phi }^2
 - (D-1) ( 2\;{\dot H} + D\,H^2 ) - 2\; {{\partial U}\over
{\partial \Phi}} - c + 2\, U  & = & 0
\label{eqcos}
\end{eqnarray}
where dot $ {}^. $ stands for $ {{d}\over{dX^0}}$, and
\begin{equation}
\rho = T_0^0  \qquad , \qquad -\delta_i^k~p = T_i^k ~.
\end{equation}
The conservation equation takes the form of eq.(\ref{cont})
\begin{equation}
\dot{\rho} + (D-1)\, H\, (p + \rho ) = 0 ~~.
\end{equation}
By defining,
\begin{eqnarray}
\Psi & \equiv & \Phi - \log{\sqrt{-G}} = \Phi - (D-1)\,\log R \cr
{\bar \rho } & =  & e^{\Phi}~\rho \quad, \quad {\bar p } =  e^{\Phi}~p ~,
\label{psiba}
\end{eqnarray}
 equations (\ref{eqcos}) can be expressed in a more compact form as
\begin{eqnarray}
{\ddot \Psi } - (D-1)\,  H^2  - \left.{{\partial U}\over
{\partial \Psi}}\right|_R & = &  {\bar \rho }  \nonumber \\ \cr
 {\dot H}  - H \, {\dot \Psi }  +  \left.{ R \over {D-1}}{{\partial U}\over
{\partial R}}\right|_{\Psi} & = &    {\bar p }  \nonumber \\ \cr
{\dot \Psi }^2 - (D-1)\,H^2  - 2\; {\bar \rho } - 2\, U  + c & = & 0
\; ,
\label{eqcof}
\end{eqnarray}
The conservation equation reads
\begin{equation}
\dot{{\bar\rho}} -  {\dot \Psi }\; {\bar \rho }+ (D-1)\, H\,  {\bar p }
= 0
\end{equation}

As is known, under the duality transformation $ R \longrightarrow
R^{-1} $ , the dilaton transforms as $\Phi \longrightarrow
\Phi + (D-1)\,\log R $. The shifted dilaton $\Psi$ defined by
eq.(\ref{psiba}) is invariant under duality.

The transformation
\begin{equation}
 R' \equiv R^{-1} \quad, ~
\end{equation}
implies
\begin{equation}
\Psi' = \Psi \quad, ~ H' = -H \quad, ~ {\bar p' }=-p \quad, ~
{\bar\rho'} = {\bar\rho}
\end{equation}
provided $ u = d $, that is, a   duality invariant  string source.
This is the duality  invariance transformation of eqs.(\ref{eqcof}).

Solutions to the effective string equations have been extensively
treated in the literature \cite{eqef} and they are not our main
purpose. For the sake of completeness, we briefly analyze the limiting
behaviour of these equations for $ R \to \infty$ and $ R \to 0 $.

It is difficult to make a complete analysis of the  effective string
equations (\ref{eqcof}) since the knowledge about the potential $ U $
is rather incomplete. For weak coupling ($ e^{\Phi} $ small ) the
supersymmetry breaking produces an effective potential that decreases
very fast (as the exponential of an exponential of $\Phi$) for
 $\Phi \to -\infty$.

Let us analyze the asymptotic behavior of  eqs.(\ref{eqcof}) for $ R
\to \infty $ and  $ R \to 0 $ assuming that the potential $ U $
can be ignored.  It is easy to see that a power behaviour Ansatz both
for $ R $ and for $ e^{\Psi} $ as functions of $ X^0 $ is consistent
with these equations. It turns out that the string sources do not
contribute to the leading behaviour here, and we find for  $ R \to 0 $
\begin{eqnarray}
R_{\mp} =&  C_1 \; (X^0)^{\pm 1/\sqrt{D-1}} ~ \to 0\quad,\cr
 e^{\Psi_{\mp}} =& C_2 \; ( X^0 )^{-1} ~ \to  \left\{
\begin{array}{ll}  \infty \\  0 \end{array} \right.
\label{solef}
\end{eqnarray}
Where $ C_1 $ and $  C_2 $ are constants.
Here the branches $(-)$ and $(+)$ correspond to $ X^0 \to 0 $ and to
 $ X^0 \to \infty $ respectively. In both regimes $ R_{\mp} \to 0 $
and $ e^{\Phi_{\mp}} \to 0$.

The potential $ U(\Phi) $ is hence negligible in these regimes. In
terms of the conformal time $ \eta $ , the behaviours (\ref{solef})
result
\begin{eqnarray}
R_{\mp} =&  C_1' \; (\eta)^{\pm {1 \over {\sqrt{D-1} \mp 1}}} \to 0 \cr
 e^{\Psi_{\mp}} =& C_2' \; (\eta)^{-{{\sqrt{D-1}}\over{\sqrt{D-1} \mp
1}}} \to  \left\{
\begin{array}{ll}  \infty \\  0 \end{array} \right.
\label{sefd}
\end{eqnarray}
Where $ C_1' $ and $  C_2' $ are constants.
The branch $(-)$ would describe an expanding non-inflationary
behaviour near the initial singularity $ X^0 = 0 $ , while the branch
$(+)$ describes a `big crunch' situation and is rather unphysical.

 Similarly,   for   $ R \to \infty $ and
 $ e^{\Phi} \to \infty $, we find
\begin{eqnarray}
R_{\mp} =&  D_1 \; (X^0)^{\mp 1/\sqrt{D-1}} ~ \to \infty \quad,\cr
 e^{\Psi_{\mp}} =& D_2 \; ( X^0 )^{-1} ~ \to  \left\{
\begin{array}{ll}  \infty \\  0 \end{array} \right.
\label{salef}
\end{eqnarray}
Where $ D_1 $ and $  D_2 $ are constants.
Here again, the branches $(-)$ and $(+)$ correspond to $ X^0 \to 0 $ and to
 $ X^0 \to \infty $ respectively, but now in both regimes
$ R_{\mp} \to \infty$ and $ e^{\Phi_{\mp}} \to \infty $. (In this
limit, one is not guaranteed that $ U $ can be consistently
neglected). In terms of the conformal time, eqs.(\ref{salef}) read
\begin{eqnarray}
R_{\mp} =&  D_1' \; (\eta)^{\mp {1 \over {\sqrt{D-1} \pm 1}}} \to \infty \cr
 e^{\Psi_{\mp}} =& D_2' \; (\eta)^{-{{\sqrt{D-1}}\over{\sqrt{D-1} \pm
1}}} \to  \left\{
\begin{array}{ll}  \infty \\  0 \end{array} \right.
\label{safd}
\end{eqnarray}
The branch $(+)$ describes a noninflationary expanding behaviour for $
X^0 \to \infty $ faster than the standard matter dominated expansion,
while the branch $(-)$ describes a super-inflationary behaviour
$\eta^{-\alpha}$, since $ 0 < \alpha < 1 $, for all D.

The behaviours (\ref{solef}) for  $ R_{\mp} \to 0 $ and (\ref{salef})
for $ R_{\mp} \to \infty$ are related by duality $ R \leftrightarrow 1/R $.

\subsection{\bf String driven inflation?}

Let us consider now the question of whether de Sitter spacetime may be
a self-consistent solution of the effective string equations
(\ref{eqcos})
with the string sources included. The strings in cosmological
universes like de Sitter spacetime have the equation of state
(\ref{rogen})-(\ref{pgen}).
Since $ e^{\Psi} = e^{\Phi} \; R^{1-D} $ :
\begin{eqnarray}
{\bar \rho} & = & e^{\Psi} \left( u \; R + {d \over R} + s \right)
\label{denb} \\ \cr
{\bar p} & = &
{{e^{\Psi}} \over {D-1}} \left( {d \over R} -  u \; R\right)
 \label{prba}
\end{eqnarray}

In the absence of dilaton potential and cosmological constant term,
the string sources do not generate de Sitter spacetime as discussed in
sec. III.A. We see that for $ U = c = 0 $ , and $ R = e^{H X^0} $ ,
eqs.(\ref{eqcof}) yields to a contradiction
 (unless $ D = 0 $ ) for
the value of $\Psi $ , required to be
 $ -H  X^0 \, + $ constant.

A self-consistent solution describing asymptotically de Sitter
spacetime self-sustained by the string equation of state
(\ref{denb})-(\ref{prba}) is given by
 \begin{eqnarray}
 R & = & e^{H X^0} ~~, ~~ H = {\rm  constant} > 0 ~~,\cr
 2U-c & = & D\; H^2
=  {\rm  constant} \cr
\Psi_{\pm} & = & \mp H X^0
\pm i\pi + \log{{(D-1)\,H^2}\over {\rho_{\pm}}} \cr
\rho_+ & \equiv & u \quad , \quad \rho_- \equiv d \quad
\label{anti}
\end{eqnarray}
The branch $\Psi_+$ describes the solution for $ R \to \infty $ ( $ X^0
\to + \infty  ) $, while the branch  $\Psi_-$ corresponds to  $ R \to 0 $
 ( $ X^0 \to - \infty  ) $. De Sitter spacetime with lorentzian
signature self-sustained by the strings necessarily requires a constant
imaginary piece $ \pm i \pi $ in the dilaton field. This makes $
e^\Psi < 0 $ telling us that the gravitational constant $ G \sim
 e^\Psi < 0 $ here describes antigravity.

Is interesting to notice that in the euclidean signature case, i. e.
(+++\ldots++), the Ansatz ${\dot H} = 0,~ 2U-c=$constant, yields a
constant curvature geometry with a real dilaton, but which is of
Anti-de Sitter type. This solution is obtained from
eqs.(\ref{prba})-(\ref{anti}) through the transformation
\begin{equation}
{\hat X}^0 = i X^0 ~~~,~~~{\hat H} = -i H  ~~~,~~~ X^i = X^i ~~~
,~~~\Psi = \Psi
\label{traf}
\end{equation}
which maps the Lorentzian de Sitter metric into the positive definite
one
\begin{equation}
d{\hat s}^2 = (d{\hat X}^0)^2 + e^{{\hat H}{\hat X}^0}~(d\vec{X})^2.
\label{posi}
\end{equation}
The equations of motion (\ref{eqcof}) within the constant curvature
Ansatz $({\dot{\hat H}} = {\ddot \Psi } = 0 )$
are mapped onto the equations
\begin{eqnarray}
 (D-1)\, {\hat H}^2  - \left.{{\partial U}\over
{\partial \Psi}}\right|_R & = &  {\bar \rho }  \nonumber \\ \cr
  {\hat H} \, {{d \Psi}\over {d {\hat X}^0 }}
+  \left.{ R \over {D-1}}{{\partial U}\over
{\partial R}}\right|_{\Psi} & = &    {\bar p }  \nonumber \\ \cr
-({{d \Psi}\over {d {\hat X}^0 }})^2
+(D-1)\,{\hat H}^2  - 2\; {\bar \rho } - 2\, U  + c & = & 0
\; ,
\label{eqhat}
\end{eqnarray}
with the solution
\begin{eqnarray}
 R & = & e^{{\hat H} {\hat X}^0} ~~, ~~ {\hat H} = {\rm  constant} > 0 ~~,\cr
 c -2 \, U & = & D\; {\hat H}^2
=  {\rm  constant} \cr
\Psi_{\pm} & = & \mp {\hat H } {\hat X}^0
+ \log{{(D-1)\,
{\hat H}^2}\over {\rho_{\pm}}} \cr
\rho_+ & \equiv & u \quad , \quad \rho_- \equiv d \quad
\label{antit}
\end{eqnarray}
Both solutions (\ref{antit}) and (\ref{anti}) are mapped one into
another through the transformation (\ref{traf}).

\bigskip

It could be recalled that in the context of (point particle) field
theory, de Sitter spacetime (as well as anti-de Sitter) emerges as an
exact selfconsistent solution of the semiclassical Einstein equations
with the back reaction included \cite{uno} - \cite{dos}. (Semiclassical in this
context, means that matter fields including the graviton are quantized
to the one-loop level and coupled to the (c-number) gravity background
through the expectation value of the energy-momentum tensor $T_A^B$ . This
expectation value is given by the trace anomaly:
$ <T_A^A> = {\bar \gamma} \; R^2  $). On the other hand,
the $\alpha'$ expansion of the effective string action admits anti-de
Sitter spacetime (but not  de Sitter) as a solution when the quadratic
curvature corrections (in terms of the Gauss-Bonnet term) to the
 Einstein action are included \cite{tres}. It appears that the
corrections to the anti-de Sitter constant curvature are qualitatively
similar in the both cases, with $\alpha'$ playing the r\^ole of the trace
anomaly parameter ${\bar \gamma}$\cite{dos}.

The fact that de Sitter inflation with true gravity $ G \sim
 e^\Psi > 0 $ does not emerge as a solution of the effective string
equations does not mean that string theory excludes inflation. What
means is that the effective string equations are not  enough to get inflation.
The effective string action is a low
energy field theory approximation to string theory containing only the
{\it massless} string modes ({\it massless} background fields).

The vacuum energy scales to start inflation (physical or true vacuum)
are typically of the order of the Planck mass \cite{romi} - \cite{linde}
where the effective string action approximation breaks
down. One must consider the massive string modes (which are absent
from the effective string action) in order to properly get the
cosmological condensate yielding de Sitter inflation. We do not have
at present the solution of such problem.


\newpage
\pagestyle{empty}
\textheight=250mm
\textwidth=165mm
\leftmargin=-15mm
\topmargin=-2cm
\begin{centerline}
{\bf Table I}

\bigskip

{\bf STRING PROPERTIES FOR ARBITRARY $R(X^0)$}

\bigskip

\end{centerline}
\vskip 20pt
\begin{tabular}{|l|l|l|l|}\hline
$ $&  $ $ & $ $ &  Equation  of State: \\
$ $& $~~$ Energy  & $~~~~$Pressure  & $ $   \\
$ $&  $ $ & $ $ & $~~~p = (\gamma-1) \rho $\\ \hline
D = 1 + 1: two families  & $ $ & $ $ & $ $ \\
\hspace{13mm} of solutions  & $ $ & $ $ & $ $\\ \hline
$ $ & $ $ & $ $ & $ $ \\
(i)$~\eta\pm X=f_{\pm}(\sigma\pm\tau)$  & $~~~E = ~ u \; R $ &
 $~~~~P = - E$ & $~~~~\gamma = 0 $ \\
 (ii)$~\eta\pm X=$~constant & $~~~E  = d /R$ &
$~~~~P = + E$ & $~~~~\gamma = 2 $ \\
$ $ & $ $ & $ $ & $ $ \\ \hline
$D = 2 + 1$: Ring Solutions,  & $ $ & $ $ & $ $ \\
three asymptotic &  $  E = {1 \over {\alpha'}}~{\dot X}^0(\tau)$
& $ P = {{R(\tau)^2} \over {2 \alpha' |{\dot X}^0(\tau)|}}\,
[{\dot f}^2 - f^2 ]  $ & $ $ \\
behaviours (u, d, s) & $ $ & $ $ & $ $ \\ \hline
$ $ & $ $ & $ $ & $ $ \\
(i) unstable for $R\to\infty$ & $E_u  \buildrel{R \to \infty}\over = u
\; R \to \infty$ & $P_u = -E/2\to-\infty$ & $~~\gamma_u = 1/2 $ \\
(ii) dual to (i) for $R \to 0$ & $E_d \buildrel{R \to 0}\over = d /R
\to \infty$ & $P_d = + E/2 \to \infty$ & $~~\gamma_d = 3/2 $ \\
(iii) stable for  $R\to\infty$ & $E_s = \;$ constant & $P_s = 0 $ &
 $~~\gamma_s = 1 $\\
$ $&  $ $ & $ $ & $ $\\ \hline
D-Dimensional spacetimes: &  $ $ & $ $ & $ $ \\
general asymptotic behaviour & $ $ & $ $ & $ $ \\ \hline
$ $ & $ $ & $ $ & $ $ \\
(i) unstable for $R\to\infty$ & $E_u  \buildrel{R \to \infty}\over = u
\; R \to \infty$ & $P_u = -{E\over{D-1}}\to-\infty$ &
$\gamma_u = (D-2)/(D-1) $ \\
(ii) dual to (i) for $R \to 0$ & $E_d \buildrel{R \to 0}\over = d /R
\to \infty$ & $P_d = + {E\over{D-1}} \to \infty$ & $\gamma_d = D/(D-1) $ \\
(iii) stable for  $R\to\infty$ & $E_s = \;$ constant & $P_s = 0 $ &
 $~~\gamma_s = 1 $\\
$ $&  $ $ & $ $ & $ $\\ \hline
\end{tabular}
\vskip 20pt
\newpage
\begin{centerline}
{\bf Table II}

\bigskip

{\bf STRING ENERGY DENSITY AND PRESSURE FOR ARBITRARY $R(X^0)$}

\bigskip

\end{centerline}
\vskip 20pt
\begin{tabular}{|l|l|l|}\hline
$ $& Energy density: $~\rho \equiv E/R^{D-1}$ & \hspace{10mm}Pressure  \\
$ $&  $ $ & $ $ \\ \hline
Qualitatively correct &  $ $ & $ $\\
formulas for all R and D &
$ ~~\rho = \left( u \; R + {{d} \over R} + s \right) {1 \over
{R^{D-1}}} $ &
$ p  = {1 \over {D-1}} \left( {d \over R} -  u \; R\right) {1 \over
{R^{D-1}}} $ \\
$ $&  $ $ & $ $ \\ \hline
\end{tabular}
\vskip 20pt
\begin{centerline}
{\bf Table III}

\bigskip

{\bf STRING COSMOLOGY IN GENERAL RELATIVITY}

\bigskip

\end{centerline}
\vskip 20pt
\begin{tabular}{|l|l|l|}\hline
Einstein equations & \hspace{4mm}Expansion factor  &\hspace{6mm} Temperature \\
(no dilaton field) & \hspace{12mm} $R(X^0)$ &  \hspace{13mm}$T(R)$ \\
\hline
$ $&  $ $ & $ $ \\
$ X^0 \to 0 $ & ${D \over 2}
\left[{{2 d}\over {(D-1)(D-2)}}\right]^{1 \over D}~ (X^0)^{2 \over D}$&
\hspace{4mm}${{dD}\over{S(D-1)}}\;1/R$\\
$ $&  $ $ & $ $ \\ \hline
$ $&  $ $ & $ $ \\
$ X^0 \to \infty$ &$\left[{{(D-2) u}\over {2(D-1)}}\right]^{1 \over
{D-2}}~ (X^0)^{2 \over {D-2}}$ &\hspace{4mm} ${{(D-2)u}\over{(D-1)S}}~R $\\
(without string decay) &  $ $ & $ $ \\ \hline
$ $&  $ $ & $ $ \\
$ X^0 \to \infty$ &$\left[{{(D-1) s}\over {2(D-2)}}\right]^{1 \over
{D-1}}~ (X^0)^{2 \over {D-1}}$&usual matter \\
(with string splitting)&  $ $ &  dominated behaviour  \\ \hline
\end{tabular}
\vskip 20pt
\newpage
\begin{centerline}
{\bf Table IV}

\bigskip

{\bf EFFECTIVE STRING EQUATIONS SOLUTIONS IN COSMOLOGY}

\bigskip

\end{centerline}
\vskip 20pt
\begin{tabular}{|l|l|l|}\hline
Effective String &  $~~R(X^0)\to 0$  &  $~~R(X^0)\to \infty$\\
\hspace{4mm}equations & \hspace{4mm}behaviour &
\hspace{4mm} behaviour \\ \hline
$ $&  $ $ & $ $ \\
\hspace{4mm}$ X^0 \to 0 $ & $~\sim\,(X^0)^{+ 1/\sqrt{D-1}}$ &
$~\sim\,(X^0)^{-1/\sqrt{D-1}}$\\
$ $&  $ $ & $ $ \\ \hline
$ $&  $ $ & $ $ \\
\hspace{4mm}$ X^0 \to \infty$ & $~\sim\,(X^0)^{-1/\sqrt{D-1}}$ &
$~\sim\,(X^0)^{+ 1/\sqrt{D-1}}$ \\
$ $&  $ $ & $ $ \\ \hline
\end{tabular}
\vskip 20pt
\begin{center}
\vskip 40pt
{\bf Table Captions}
\end{center}
\vskip 12pt
Table I. String energy and pressure as obtained from exact string
solutions for various expansion factors  $R(X^0)$.
\vskip 36pt
Table II. The string energy density and pressure for a gas of strings
can be summarized by the above formulas which become exact for $R \to
0$ and for $R \to \infty$.
\vskip 36pt
Table III. The {\bf self-consistent} cosmological solution of the
Einstein equations in General Relativity with the string gas as
source.
\vskip 36pt
Table IV. Asymptotic solution of the string effective equations
(including the dilaton).
\end{document}